\documentclass[11pt,aps,reprint,citeautoscript,superscriptaddress,floatfix]{revtex4-2}

\usepackage[utf8]{inputenc}

\usepackage[margin={2cm,3cm}]{geometry}

\usepackage{graphicx}
\usepackage{physics}
\usepackage[labelsep=period,tableposition=top]{caption}
\usepackage{subcaption}
\usepackage{tabularx}
\usepackage{makecell}
\usepackage{multirow}
\usepackage{physics}

\begin{document}


\title{Temperature dependence of the AB-lines and Optical Properties of the Carbon-Antisite Vacancy Pair in 4H-SiC}
\author{Oscar Bulancea-Lindvall}%
\author{Joel Davidsson}%
\author{Ivan G. Ivanov}%
\affiliation{Department of Physics, Chemistry and Biology (IFM), Linköping University, SE-58183 Linköping, Sweden}
\author{Adam Gali}%
\affiliation{HUN-REN Wigner Research Centre for Physics, P.O. Box 49, H-1525 Budapest, Hungary}
\affiliation{Department of Atomic Physics, Institute of Physics, Budapest University of Technology and Economics, Műegyetem rakpart 3., H-1111 Budapest, Hungary}
\affiliation{MTA-WFK Lend\"ulet "Momentum" Semiconductor Nanostructures Research Group}
\author{Viktor Ivády}%
\affiliation{Department of Physics, Chemistry and Biology (IFM), Linköping University, SE-58183 Linköping, Sweden}
\affiliation{Department of Physics of Complex Systems, E\"otv\"os Loránd University, Egyetem tér 1-3, H-1053, Budapest, Hungary}
\affiliation{MTA–ELTE Lend\"{u}let "Momentum" NewQubit Research Group, Pázmány Péter, Sétány 1/A, 1117 Budapest, Hungary}
\author{Rickard Armiento}%
\author{Igor A. Abrikosov}%
\affiliation{Department of Physics, Chemistry and Biology (IFM), Linköping University, SE-58183 Linköping, Sweden}

\begin{abstract}
\noindent
Defects in semiconductors have in recent years been revealed to have interesting properties in the venture towards quantum technologies. In this regard, silicon carbide has shown great promise as a host for quantum defects. In particular, the ultra-bright AB photoluminescence lines in 4H-SiC are observable at room temperature and have been proposed as a single-photon quantum emitter. These lines have been previously studied and assigned to the carbon antisite-vacancy pair (CAV). In this paper, we report on new measurements of the AB-lines' temperature dependence, and carry out an in-depth computational study on the optical properties of the CAV defect. We find that the CAV defect has the potential to exhibit several different zero-phonon luminescences with emissions in the near-infrared telecom band, in its neutral and positive charge states. However, our measurements show that the AB-lines only consist of three non-thermally activated lines instead of the previously reported four lines, meanwhile our calculations on the CAV defect are unable to find optical transitions in full agreement with the AB-line assignment. In the light of our results, the identification of the AB-lines and the associated room temperature emission require further study.
\end{abstract}

\maketitle

%



\section{Introduction}
Quantum devices for information processing and communication will require fundamental components that are robust against noise and practical for efficient state manipulation and photon emission. Point defects qubits found in wide bandgap semiconductors are promising candidates for such applications because of their substantial display of optical and spin properties, and their potential for localized spin states well isolated from decoherence sources \cite{Weber2010,Koehl2011,Ladd2010,Doherty2013,Yan2018,Son2020,Castelletto2020,Chuang2013}. Their capacity for single-photon emission is also a vital ability for quantum information applications, e.g., communication and cryptography \cite{Brassard2000}. 

The nitrogen-vacancy (NV) center in diamond has been leading the field among defects studied in the last decade, with demonstrated coherent control of its spin states and bright single-photon emission even at ambient conditions \cite{Doherty2013,Kurtsiefer2000,Stanwix2010}. 
The success of this defect in various quantum device applications has invigorated the search for defects with similar or improved properties, including the the consideration of new hosts with potential technological advantages.
In particular, silicon carbide (SiC) which shares many of the relevant properties of diamond and can be found in a myriad of polytypes with capabilities to host numerous color centers, is regarded as a promising alternative host \cite{Weber2010,Falk2013,Castelletto2020,CubicSiCPaper,Lohrmann2017}. Its established presence in industry has led to mature high-power electronics and advanced device fabrication techniques, which could prove to be an advantage in upcoming spin-photonic applications \cite{Yamada2011,Lukin2020}.

Recent studies have showcased a number of interesting defects in SiC with promising abilities as near-telecom wavelength emitters and qubits with coherent lifetimes rivaling the NV center \cite{Szasz2015,Koehl2011,Castelletto2020,Magnusson2018,Miao2020}. Furthermore, measurements of the 4H- and 6H-SiC polytypes have shown that these hosts are the source of ultra-bright photoluminescence (PL) \cite{Steeds2009}, in the range of 640--680 nm for 4H-SiC, emitting at a rate of Mcps \cite{Castelletto2014}. Such a high brightness is rare for defects in bulk materials and has the potential to become a vital component in quantum optics with the significant enhancement of photonic nanocavities \cite{Lukin2020,Yamada2011}.

Observations of this emission at nitrogen temperature (80 K) have found that it consists of eight lines, dubbed A1--A4 and B1--B4 in order of increasing wavelength. Six of these lines (A1, A2, B1--B4) were studied and initially assigned to the neutral carbon antisite-vacancy pair (CAV) in early works based on the available optical measurements \cite{Steeds2002,Steeds2009}. The two remaining lines, A3 and A4, were later presented in the spectrum at 80 K in Ref.~\onlinecite{Castelletto2014}, which also reported \emph{ab initio} calculations finding no visible-light photoluminescence for the neutral CAV state. Ref.~\onlinecite{Castelletto2014} instead provided a reasonable model for how the eight lines could be assigned to the four configurations of the positively charged CAV defect (CAV${}^+$). The A2, A4, B2, and B4 were attributed to zero-phonon lines of the four nonequivalent defect configurations in 4H-SiC, whereas the A1, A3, B1, and B3 lines were suggested to originate from respective second split-off excited states, emerging due to a Jahn-Teller splitting of degenerate energy level pairs in the excited state, which are temperature-activated at 80 K \cite{Castelletto2014}. The spectrum was therefore reinterpreted as belonging to the positive charge state, pending a complete characterization of its excited state dynamics.

The assignment of the AB-lines to the CAV${}^+$ defect has since been prominent in the field, leading to its use in studies of defect formation and fabrication. In particular, the AB-lines have been used as markers for CAV defect formation and incorporated into models for defect conversion pathways \cite{karsthofConversionPathwaysPrimary2020}. The assignment has similarly been applied to ascertain the efficiency of CAV${}^+$ creation in micropillar fabrication and charge control via Schottky barrier diodes \cite{bathenChargeStateControl2023}.  

However, the spectra observed at 7 K in Ref.~\onlinecite{Steeds2009} display significant contributions from the B1 and B3 lines. This contribution should be weaker at 7 K in view of the also observed respective 4 meV separation from the B2 and B4 lines, suggesting that the observed contribution may be caused by local laser heating. Furthermore, while the temperature dependence for the B-lines has been presented \cite{Steeds2009}, the behavior of the A-lines has not been investigated.

In the present work, we report on the complete temperature dependence of the spectrum which compels us to revisit the assignment of these lines to the CAV defect.

In addition, despite the extensive knowledge of the structure and ground state properties of the CAV defect showcased in previous works and the motivations behind the AB-line identification \cite{Castelletto2014}, its excited states and spin properties have yet to be explored by \textit{ab-initio} methods, taking into account all charge states and configurations. A thorough characterization for the CAV could contribute to the working model for the origin of the AB-lines and reveal additional properties as a quantum emitter and qubit. Modern first principle methods have proven their ability to provide accurate predictions of material properties, in particular for optical and spin properties of defects \cite{Deak2010}. Therefore, this paper will also present a detailed first principle characterization of the CAV defect, involving its optical and spin-related properties relevant for the previously highlighted areas of interest and provide data for comparison to the measurements on the AB-lines at low temperatures. In particular, beyond what has previously been characterized, we give estimates for radiative lifetimes and polarizations of all calculated zero-phonon lines (ZPLs). In addition, we solve the Bethe-Salpeter equation \cite{rohlfingElectronholeExcitationsOptical2000} on top of a single-shot Green's functions-based $GW$ approximation ($G0W0$) \cite{hedinNewMethodCalculating1965} to the CAV${}^+$ system, which has previously only been considered for the neutral CAV state \cite{Szasz2015}.

This paper is structured as follows. In section \ref{PL}, we present the experimental setup and results of photoluminescence measurements on the AB-lines. Then, in section \ref{sec:first principles}, we provide a description and introduction to the CAV defect in section \ref{sec:CAV background}, detail the first principle characterization, starting with an introduction to the examined quantities and tools in section \ref{sec:methods}, and present relevant results in section \ref{sec:results}. Finally, in section \ref{sec:Discussion}, we reflect on the relevance of the defect as a quantum emitter and its relation to the AB-lines.

\section{Photoluminescence in the AB-line Spectrum}\label{PL}
For the measurement of the AB-lines at low temperatures, we use commercial high-purity semi-insulating (HPSI) 4H-SiC substrate that has been irradiated with 2 MeV electrons to a fluence of $10^{17}$ cm\textsuperscript{$-$2}. The substrate has the crystal c-axis nearly perpendicular to the surface. The excitation laser wavelength used to obtain the photoluminescence (PL) spectra of the AB-lines is 514.5 nm. The laser ($\sim$15 mW in power) is moderately focused to a spot of $\sim$2 mm in diameter at the sample to avoid local heating. The PL is excited through and collected from the edge of the sample, to ensure that light with both parallel and perpendicular polarization to the c-axis can be registered. The temperature dependence of the lines is measured in a liquid-He operated cryostat with a temperature controller. The PL was registered using a double monochromator (SPEX 1404) equipped with GaAs-photocathode photomultiplier.
\begin{figure}[h]
    \centering
    \includegraphics[width=0.9\columnwidth]{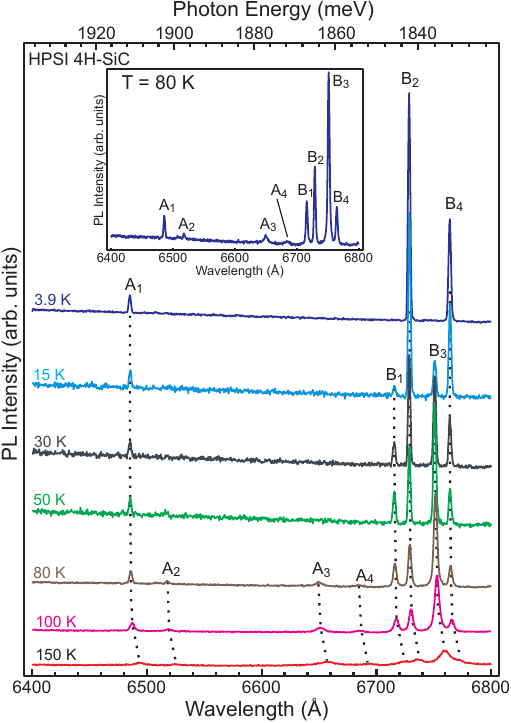}
    \caption{Temperature dependence of the AB-lines in a HPSI 4H-SiC substrate obtained with 514.5 nm excitation under low power density conditions ($<$ 0.5 W/cm\textsuperscript{2}). The insert focuses on the spectrum at 80 K, which is akin to the spectrum in Fig. 1a of Ref.~\onlinecite{Castelletto2014}. Notice that the low-temperature spectrum at 3.9 K (top curve) only exhibits the A1, B2 and B4 lines, whereas the A2, A3, A4 lines, similar to B1 and B3, only appear at higher temperatures and are most probably associated with higher excited states of the defect responsible for the B1--B4 PL lines.%
    }
    \label{fig:AB_figure}
\end{figure}

The AB-line spectrum is obtained at temperatures between 3.9 K and 150 K (see Fig. \ref{fig:AB_figure}), including the 80 K point which is presented in Ref.~\onlinecite{Castelletto2014}. In the spectrum at 80 K (cf. the inset in Fig. \ref{fig:AB_figure}), we can clearly see the eight lines A1--A4 and B1--B4 shown in Ref.~\onlinecite{Castelletto2014} that have been proposed to be due to localized excitations in the positive charge state of the CAV \cite{Castelletto2014,Castelletto2015}. In fact, the spectrum is nearly identical to the one presented in Ref.~\onlinecite{Castelletto2014}.

The model of the CAV${}^+$ defect predicts four PL lines, A2, A4, B2 and B4, originating from transitions to lowest energy excited states in each of the four CAV${}^+$ configurations in 4H-SiC. The remaining four lines, A1, A3, B1 and B3 are not observable at low (liquid helium, LHe) temperatures because the higher-energy counterparts of the excited states, split off by the Jahn-Teller effect, are not populated. Hence, the model predicts that the latter four lines will emerge at higher temperatures when the higher-energy excited states become significantly populated (e.g., at 80 K used for the spectrum in Ref.~\onlinecite{Castelletto2014}). The above notion can easily be checked if the temperature dependence of the spectrum is measured down to LHe temperature, as displayed in Fig. \ref{fig:AB_figure}.

We notice that the experimental data is in conflict with the model suggested in Ref.~\onlinecite{Castelletto2014}. Firstly, B1 and B3 indeed vanish at LHe temperature, and are thus associated with higher-energy excited states of the defects producing the ZPLs B2 and B4, respectively. However, the A2 and A4 lines also vanish, indicating that they cannot be ZPLs corresponding to transitions between a ground state and corresponding lowest energy excited state. Secondly, the A1 line is not a higher-energy counterpart of the A2 line because it does not vanish at LHe temperature. Thus, the low-temperature spectrum of the AB-lines displays only three lines, A1, B2 and B4, instead of four corresponding to the number of CAV${}^+$ configurations in 4H-SiC. Hence, the experimental data leads us to conclude that the CAV${}^+$ model proposed in Ref.~\onlinecite{Castelletto2014} does not correspond to the observed AB-line spectrum.

We notice, however, that our results still support the notion that the lines are due to the same defect in different configurations. This can be asserted because of the intensity ratios between A1, B2, and B4 being consistently equal, which we have also observed in various other samples at low temperature (2--4 K). 

\section{First principles characterization}\label{sec:first principles}
\subsection{The carbon-antisite vacancy pair}\label{sec:CAV background}
\begin{figure}[h]
    \centering
    \includegraphics[width=0.9\columnwidth]{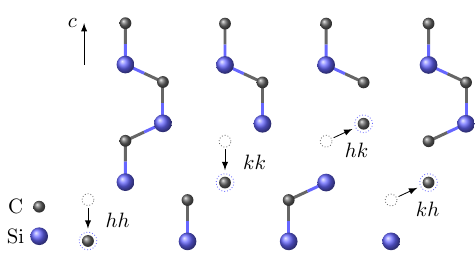}
    \caption{Illustrations of the vacancy and carbon\--antisite placement for the possible configurations of the CAV defect in 4H-SiC.}
    \label{fig:cav illustration}
\end{figure}

The carbon vacancy antisite pair is an intrinsic defect in SiC, which may exist in four non-equivalent configurations, as illustrated in Fig. \ref{fig:cav illustration}. It is stoichiometrically equivalent to the silicon vacancy, as they can be interchanged by the migration of a nearest neighbor carbon of the silicon vacancy. This transformation is predicted to have an activation barrier of 1.9--2.7 eV and can be effectuated by annealing at temperatures of approximately 750${}^\circ$C \cite{migration,Rauls2000}.  
Two of its configurations have the carbon displacement axis parallel to the crystal c-axis, which are usually denoted $hh$ and $kk$ because of the local hexagonal (h) and cubic (k) environment of both the carbon vacancy and the carbon antisite. These on-axis configurations exhibit $C_{3v}$ symmetry, resulting in the appearance of an $a_1$ and a twice degenerate $e$ one-electron state in the band gap, similar to that of the NV-center in diamond \cite{Gali2019}. The other configurations, denoted $hk$ and $kh$ (cf. Fig. \ref{fig:cav illustration}), are off-axis, meaning that they do not share the axial symmetry of the unit cell and therefore exhibit $C_{1h}$ symmetry. 

The effect of this lowered symmetry in the band structure can be understood as a splitting of the degenerate $e$-state into corresponding $a'$ and $a''$ states. 
Initial structural characterization identified the single positively charged and neutral state to be $C_{3v}$ symmetric with an occupied local $a_1$ state and an unoccupied $e$ state inside the bandgap \cite{Umeda2006,Umeda2007}. However, following \textit{ab-initio} studies found there to be a Jahn-Teller (JT) distortion taking place upon occupation of one of the $e$ states, affecting the symmetry in certain charge states \cite{Szasz2015}. Specifically, the neutral charge state in the axial configurations were suggested to be more stable in a $C_{1h}$ symmetrical configuration due to the JT effect. Similar observations were made for the negatively charged state, also occupying one of the otherwise degenerate $e$-bands \cite{Umeda2006}. This implies the need for special care in characterizing level transitions, as a change in occupation could also accompany structural and behavioral changes in the electronic structure through the introduction or loss of the JT distortion. The $hk$ and $kh$ would not require such special treatment as these configurations already split the degenerate bands owing to their natural \textit{per se} $C_{1h}$ symmetry.

\subsection{Computational methods}\label{sec:methods}
In this work, we calculate excited state properties within the Kohn-Sham density functional theory, mainly using the method of constrained occupations \cite{Gali2009a}. To aid the identification of the defect-localized states, we also employ the measure known as the inverse participation ratio (IPR) \cite{IPR_ref} of a wavefunction $\phi$,
\begin{equation}
    \operatorname{IPR}(\phi) = \frac{\int_V |\phi|^4 d^3\vec{r}}{\left(\int_V |\phi|^2 d^3\vec{r}\right)^2},
\end{equation}
where $V$ denotes the supercell volume.
This measure yields values within the interval $[1/V, 1]$, where its extreme values of $1/V$ and $1$ are given for the special cases of a fully delocalized and a point-localized wavefunction, respectively, giving a reliable measure of the localization within these bounds.

Among the quantities we include in the optical characterization, we present the zero-phonon lines (ZPLs) of all characterized transitions, corresponding transitional dipole moments, and radiative lifetimes. ZPLs are determined by the $\Delta$SCF-methodology \cite{galiTheorySpinConservingExcitation2009}, comparing the total energies of the states
\begin{equation}\label{eq:ZPL}
    E_{\text{ZPL}} = E_{\text{ex,tot}} - E_{\text{ground,tot}} \thinspace +E_{\text{corr}},
\end{equation}
using the structurally relaxed total excited state energy $E_{\text{ex,tot}}$ and ground state energy $E_{\text{ground,tot}}$, with a possible correction, $E_\text{corr}$, due to finite-size and charge-interaction effects.

The two terms on the r.h.s.\ of (\ref{eq:ZPL}) provides an accurate description for transitions between localized defect states, when finite-size effects are effectively cancelled out. 
Using the Heyd, Scuseria, and Ernzerhof exchange correlation functional (HSE06)  \cite{Heyd2003}, results have also been shown to agree with experimental measurements up to an error margin of approximately 0.1 eV \cite{Deak2010}.

However, transitions involving bound excitons in charged defect states require extra caution as the excited electron state may be particularly delocalized from the defect site. In these cases, an additional charge correction needs to be provided, to account for the effect of the charge jellium background in addition to the considerable delocalization of electron state.

Popular model-based charge correction schemes such as those by Freysoldt–Neugebauer–Van de Walle  \cite{FNV_correction}, Makov-Payne  \cite{MPPaper}, and Lany-Zunger \cite{LZ_correction} assume a charge model based on a localized defect charge density and are not likely to accurately incorporate the exciton in their descriptions. Describing an extended electron state with a local charge model could possibly be achieved by choosing a large effective Bohr-radius, or by adjusting the charge of the center to fit the supercell size scaling result. Both approaches are difficult to apply accurately without detailed knowledge of the scaling beforehand. For this reason, we instead investigate the supercell convergence behavior in order to estimate the size of the $E_{\text{corr}}$ term. We limit this analysis to the case of the positive charge state of the $hh$ configuration, under the assumption that the correction is largely independent of the actual atomic configuration of the CAV defect.

We estimate the radiative lifetime of the transitions according to\cite{stoneham}
\begin{equation}
    \tau_{f,i} = \frac{3\epsilon_0 h^4 c^3}{2(2\pi)^3n E_{f,i}^3|\vec{\mu}_{f,i}|^2},
\end{equation}
where $E_{f,i}$ is the energy difference between the compared states, taken as $E_{\text{ZPL}}$ in this work, $n$ being the refractive index of 4H-SiC, here chosen as $n=2.6473$, and transitional dipole moments, 
\begin{equation}
    \vec{\mu}_{f,i} = \bra{\phi_f}e\vec{r}\ket{\phi_i},
\end{equation}
with $\phi_f$ and $\phi_i$ being the final and initial electronic wavefunctions of the transition. While this element should ideally be evaluated using the many-body wavefunctions of the defect states, they are here approximated by the Kohn-Sham orbitals involved in the transition. Such a simplification is non-trivial, but has been shown for native defects in SiC to reproduce polarizations observed in experiments and provide quantitative agreement with measured lifetimes, with optimal accuracy for the case of using final and initial states taken from the relaxed geometries of respective excited and ground states\cite{Davidsson2020}. 

As an alternative approach to the above methods of optical characterization, we additionally apply single-shot $GW$ calculations ($G_0W_0$)\cite{hedinNewMethodCalculating1965}, adding the quasi-particle correction to the defect levels, and solve the Bethe-Salpeter equation (BSE)\cite{rohlfingElectronholeExcitationsOptical2000} to more accurately describe the screening in the electron-hole interaction of bound-to-free transitions. With these methods, we obtain the dielectric function of the relevant CAV states which we relate to the absorption spectrum. This technique has previously been applied to the carbon vacancy in SiC to include strong excitonic effects in the photo-ionization process, demonstrating an ability to predict excitation thresholds\cite{Bockstedte2010}.

In the first principle characterization, we employ highly accurate DFT calculations as implemented in the Vienna Ab-initio Simulation Package (VASP) \cite{VASP1,VASP2} using the projector augmented wave method (PAW). Calculations are performed in a 576 atom supercell containing a single CAV center, using the hybrid HSE06 functional \cite{Heyd2003} at the $\Gamma$-point, with a 420 eV plane wave cutoff. For high convergence, a $10^{-6}$ and $5\cdot 10^{-5}$ eV limit is applied for the respective density and structural relaxations. For practical considerations of the constrained occupation calculations in VASP, see the Appendix.

\begin{figure*}[t!]
    \begin{subfigure}{0.455\columnwidth}
        \centering
        \includegraphics[height=5cm,trim=0.33cm 0 0 0, clip]{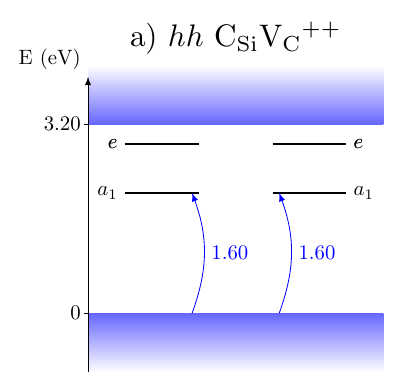}
        \label{fig:hh_cp2_result}
    \end{subfigure}
    \qquad
    \begin{subfigure}{0.47\columnwidth}
        \centering
        \includegraphics[height=5cm]{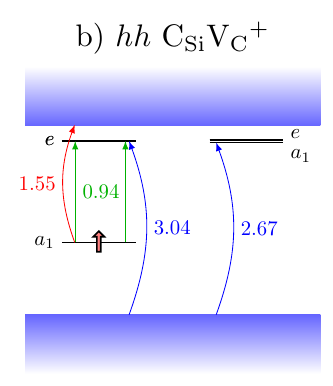}
        \label{fig:hh_cp_result}
    \end{subfigure}
    \begin{subfigure}{0.47\columnwidth}
        \centering
        \includegraphics[height=5cm]{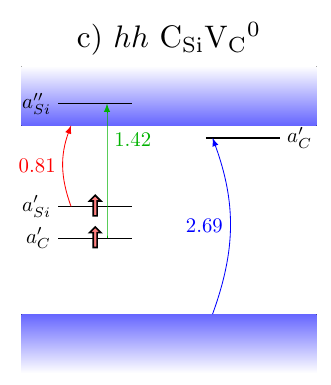}
        \label{fig:hh_c0_result}
    \end{subfigure}
    \begin{subfigure}{0.47\columnwidth}
        \centering
        \includegraphics[height=5cm]{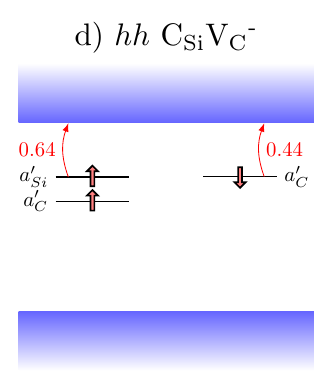}
        \label{fig:hh_cm1_result}
    \end{subfigure}
    
    \begin{subfigure}{0.455\columnwidth}
        \centering
        \includegraphics[height=5cm,trim=0.33cm 0 0 0, clip]{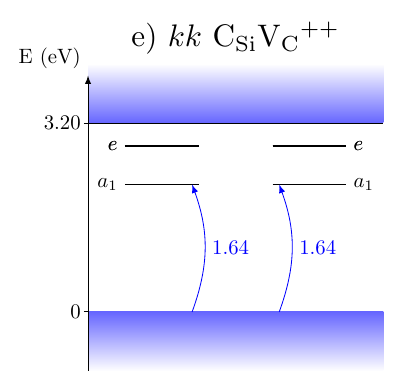}
        \label{fig:kk_cp2_result}
    \end{subfigure}
    \qquad
    \begin{subfigure}{0.47\columnwidth}
        \centering
        \includegraphics[height=5cm]{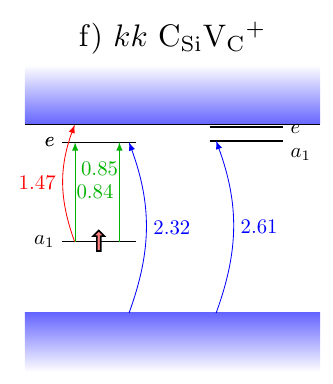}
        \label{fig:kk_cp_result}
    \end{subfigure}
    \begin{subfigure}{0.47\columnwidth}
        \centering
        \includegraphics[height=5cm]{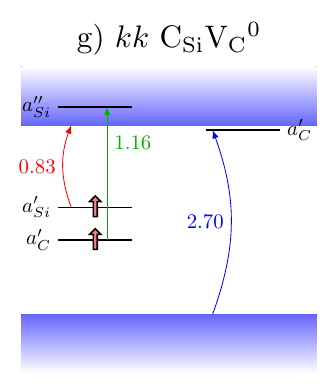}
        \label{fig:kk_c0_result}
    \end{subfigure}
    \begin{subfigure}{0.47\columnwidth}
        \centering
        \includegraphics[height=5cm]{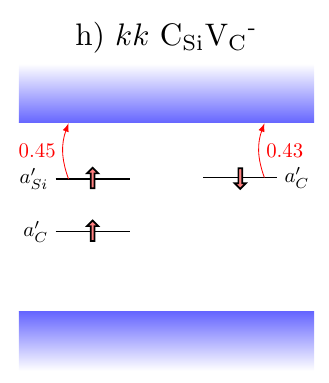}
        \label{fig:kk_cm1_result}
    \end{subfigure}
    \caption{The band structure and ZPL energies of respective possible transition for the CAV $hh$ and $kk$ configurations, showing the double positive (a and e), positive (b and f), neutral (c and g) and negative (d and h) charge states. ZPL values are displayed in eV, with blue arrows free-to-bround, red arrows bound-to-free and green arrows fully localized transitions. One should note that the positive states display two local transitions, varying in energy due to the JT relaxation which occurs after excitation onto the degenerate states. }
    \label{fig:hh_bands}
\end{figure*}

Hyperfine and zero-field splitting (ZFS) parameters are obtained under the same settings but using $\Gamma$-point PBE Kohn-Sham orbitals for the ZFS. This is done as a post-processing step using the methods presented in Ref.~\onlinecite{HyperfinePaper} and Ref.~\onlinecite{ZFSpaper} as implemented in VASP.

$G_0W_0$ VASP calculations were performed on top of HSE06-converged structure and wavefunctions of a 256 atom supercell, using more than ten times as many unoccupied bands as occupied and applying a $GW$ energy cutoff of 200 eV. The BSE calculation is done on top of the $GW$ solution, including 100 electron-hole pairs, which by the resulting $GW$ quasiparticle energies should account for transitions up to 4 eV above gap.

\subsection{Characterization results}\label{sec:results}
\begin{figure*}[t!]
    \begin{subfigure}{0.485\columnwidth}
        \centering
        \includegraphics[height=5cm]{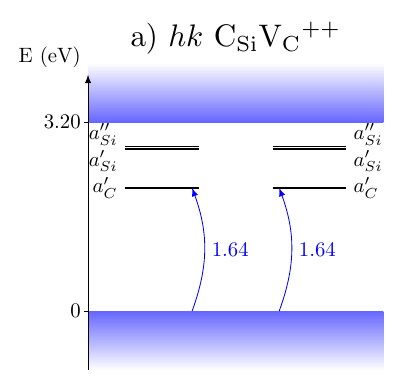}
        \label{fig:hk_cp2_result}
    \end{subfigure}
    \qquad
    \begin{subfigure}{0.47\columnwidth}
        \centering
        \includegraphics[height=5cm]{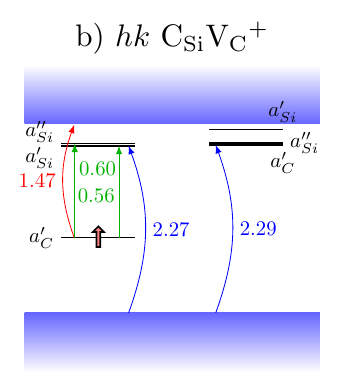}
        \label{fig:hk_cp_result}
    \end{subfigure}
    \begin{subfigure}{0.47\columnwidth}
        \centering
        \includegraphics[height=5cm]{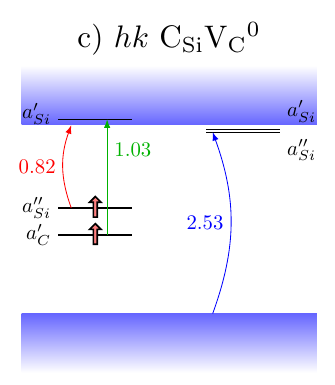}
        \label{fig:hk_c0_result}
    \end{subfigure}
    \begin{subfigure}{0.47\columnwidth}
        \centering
        \includegraphics[height=5cm]{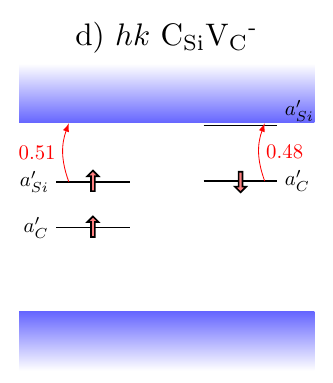}
        \label{fig:hk_cm1_result}
    \end{subfigure}
    
    \begin{subfigure}{0.485\columnwidth}
        \centering
        \includegraphics[height=5cm]{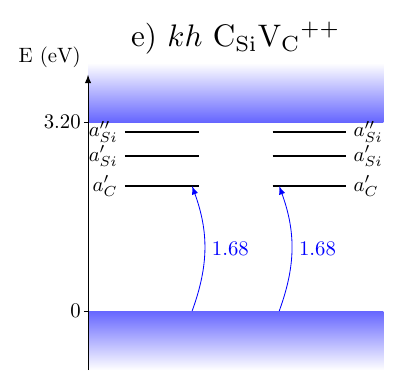}
        \label{fig:kh_cp2_result}
    \end{subfigure}
    \qquad
    \begin{subfigure}{0.47\columnwidth}
        \centering
        \includegraphics[height=5cm]{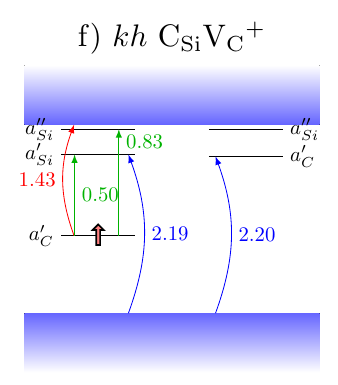}
        \label{fig:kh_cp_result}
    \end{subfigure}
    \begin{subfigure}{0.47\columnwidth}
        \centering
        \includegraphics[height=5cm]{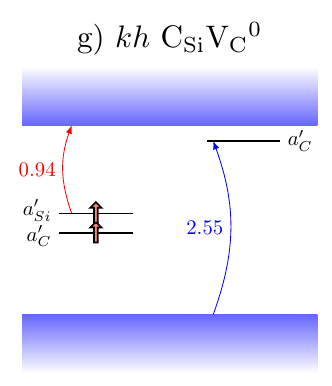}
        \label{fig:kh_c0_result}
    \end{subfigure}
    \begin{subfigure}{0.47\columnwidth}
        \centering
        \includegraphics[height=5cm]{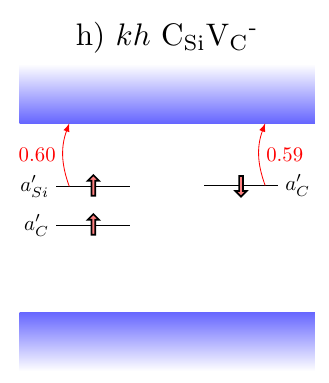}
        \label{fig:kh_cm1_result}
    \end{subfigure}
    \caption{The band structure and ZPL energies of respective possible transition for the CAV $hk$ and $kh$ configurations, showing the double positive (a and e), positive (b and f), neutral (c and g) and negative (d and h) charge states. ZPL values are displayed in eV, with blue arrows free-to-bround, red arrows bound-to-free and green arrows fully localized transitions. The reader should note the observable splitting in the positive state, yielding varying differences in ZPLs between the transitions $a_C' \to a_{Si}'$ and $a_C' \to a_{Si}''$ on the size of 0.04 and 0.3 eV for the respective $hk$ and $kh$ cases. }
    \label{fig:hk_bands}
\end{figure*}

The defect ground states and their respective Kohn-Sham band structures are shown in Figs. \ref{fig:hh_bands} and \ref{fig:hk_bands} for the on-axis $hh$ and $kk$, and the off-axis $hk$ and $kh$ configurations, respectively. As presented, the charge states considered in this study are the neutral (CAV\textsuperscript{0}) state in a $S=1$ triplet state, the positive (CAV\textsuperscript{+}) in a $S=1/2$ spin state, the double positive (CAV\textsuperscript{++}) with $S=0$, and the negative (CAV\textsuperscript{--}) state with $S=1/2$. However, the neutral ground state may in fact form either a triplet or a spin singlet state, which respectively exhibit $C_{1h}$ and $C_{3v}$ structural symmetry. In this work, the singlet state is not considered for an optical characterization, as we have found the triplet state to be more energetically favorable, in agreement with other works \cite{Szasz2015}, with an energy separation of ${\sim}0.3$ eV. Alternative spin states for the other charge states are not found in this study.

Out of the four dangling bonds of the CAV, one is found to be deep in the valence band (not shown in Figs. \ref{fig:hh_bands} and \ref{fig:hk_bands}) roughly 0.9 eV below the valence band edge, while the other three states are typically found inside the band gap. This varies among the various charge states, as the Jahn-Teller splitting may cause the highest lying dangling bond to rise into the conduction band. This is true for the neutral and negative charge states, for which the two defect states that end up in the band gap display $a'$ character, with the exception of the $hk$ neutral charge state where the higher lying state in the gap is of $a''$ representation. In charge states where the JT splitting does not take place, the states are found within the gap with $a_1$ and degenerate $e$ state characters where $C_{3v}$ symmetry applies.

\begin{figure}[h!]
    \centering
    \begin{subfigure}{0.3\columnwidth}
        \includegraphics[trim=3cm 0 5cm 0, clip,width=\columnwidth]{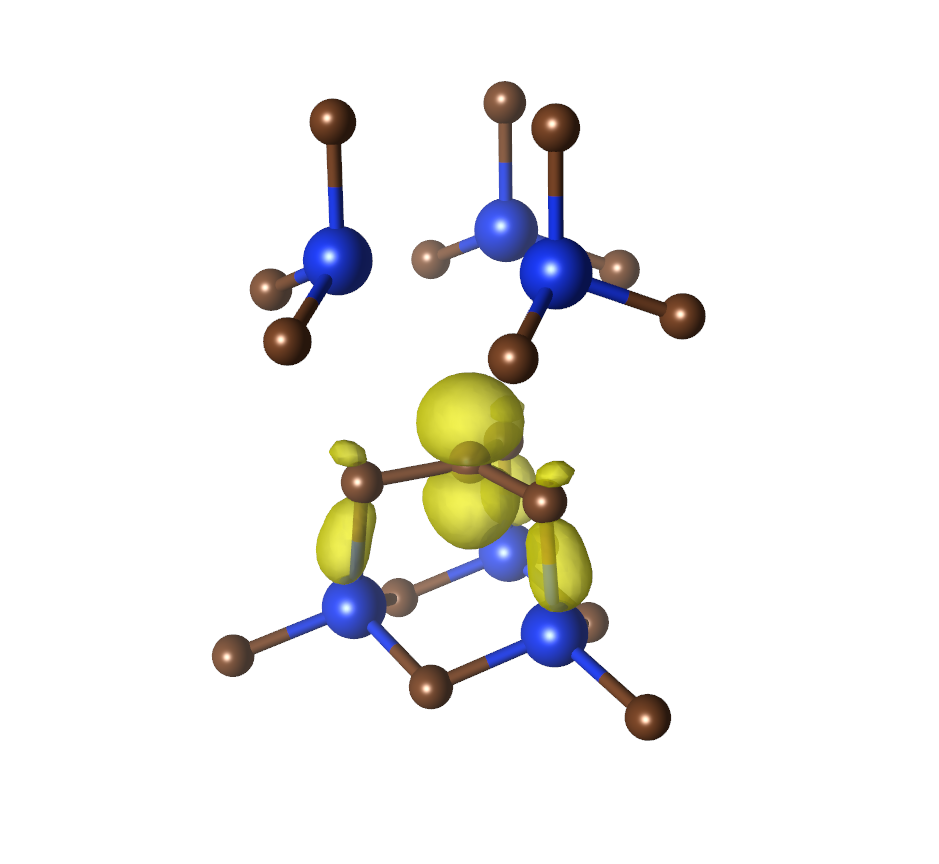}
        \caption{}
         \label{fig:charge_density1}
    \end{subfigure}
    \begin{subfigure}{0.3\columnwidth}
        \includegraphics[trim=3cm 0 5cm 0, clip, width=\textwidth]{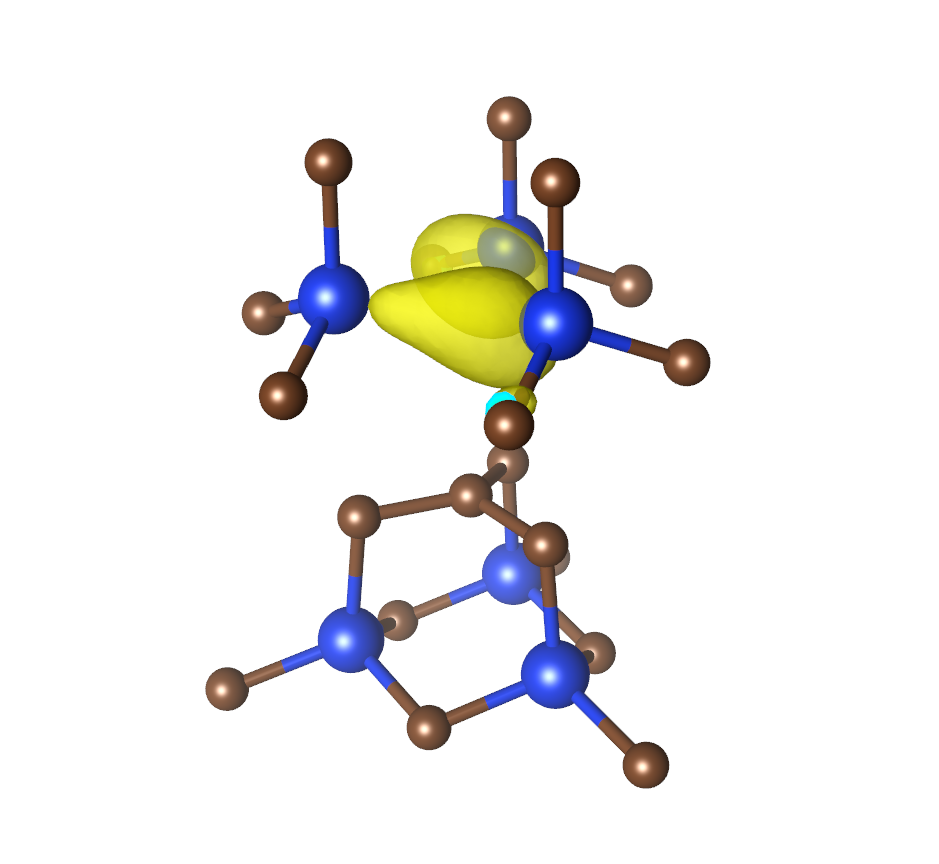}
        \caption{}
        \label{fig:charge_density2}
    \end{subfigure}
    \begin{subfigure}{0.3\columnwidth}
        \includegraphics[trim=3cm 0 5cm 0, clip,width=\columnwidth]{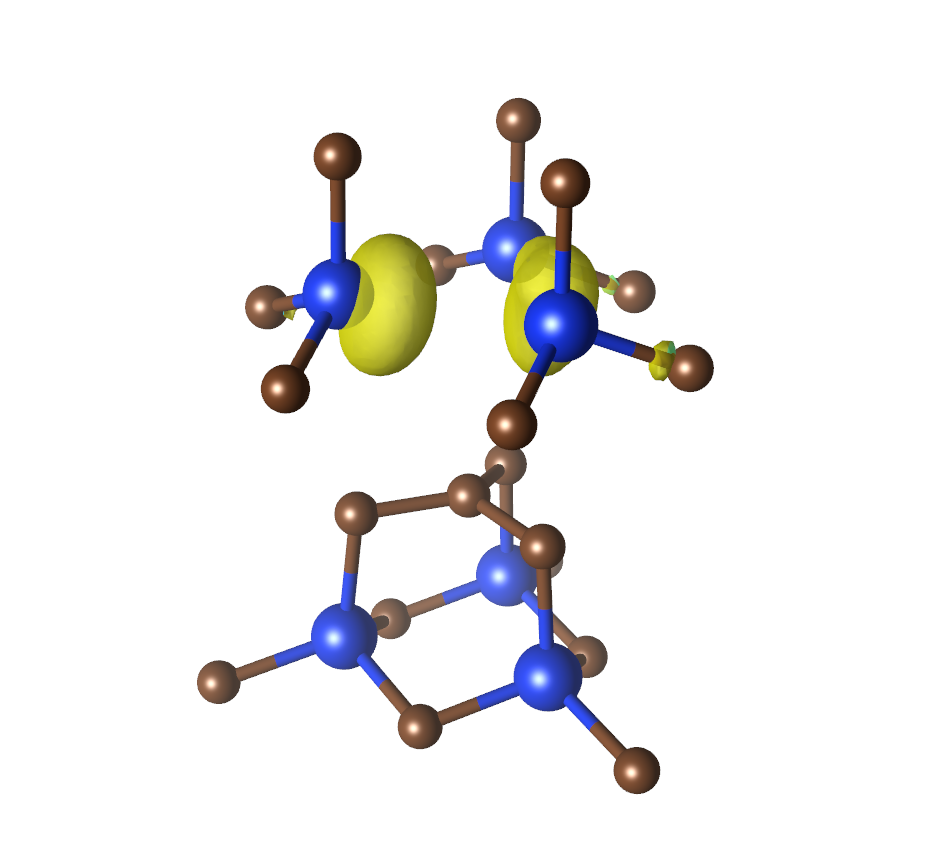}
        \caption{}
        \label{fig:charge_density3}
    \end{subfigure}
    \caption{Defect Kohn-Sham states of the $hh$ CAV${}^+$, with the $a_1 (a_C')$ state (a) and the two $e (a_{Si}')$ and $e (a_{Si}'')$ states (b and c respectively). Plotted with the package for Visualization for Electronic and Structural Analysis (VESTA) \cite{VESTA}. }
    \label{fig:charge_densities}
\end{figure}

Among these three states, the $a_1$ ($a'$) state, which is in all cases lower lying, is more localized on the antisite carbon atom (see Fig. \ref{fig:charge_density1}) and the $e$ ($a'$ and $a''$) states are gathered in proximity to the neighboring silicon atoms of the vacant carbon (see Figs. \ref{fig:charge_density2} and \ref{fig:charge_density3}). To provide a unique identifier to these states in the lower symmetry cases, the carbon-localized state will hereafter be denoted by $a_C'$ and the others by $a_{Si}'$ and $a_{Si}''$.

\begin{table*}[ht!]
    \renewcommand{\arraystretch}{1.3}
    \centering
    \caption{Properties of the studied excitations in the CAV${}^+$ state, showing ZPL, radiative lifetime and dipole moment as calculated from first principles, categorized by the exhibiting ground state. Transitions are labeled by the resulting hole and occupied state, discerning between the two $a'$ states formed from the $a_1$ and $e$ states under symmetry breaking by the localization around the carbon ($a_1 = a_C'$) or the vacancy-neighboring silicon ($e$, or $a_{Si}'$ and $a_{Si}''$).}
    
    \begin{ruledtabular}
\begin{tabular}{ccccc|ccccc}
\multicolumn{5}{c}{$\mathbf{hh}$} & \multicolumn{5}{c}{$\mathbf{kk}$}\\
Transition & \makecell{ZPL \\ (eV)} & \makecell{Lifetime \\ ($\mu$s)} & \makecell{$|\vec{\mu}_{xy}|$ \\ (Debye)} & \makecell{$|\vec{\mu}_{z}|$ \\ (Debye)} & Transition & \makecell{ZPL \\ (eV)} & \makecell{Lifetime \\ ($\mu$s)} & \makecell{$|\vec{\mu}_{xy}|$ \\ (Debye)} & \makecell{$|\vec{\mu}_{z}|$ \\ (Debye)}\\
\cline{1-10}
$a_1\to \text{CBM}$ & 1.55 & 0.32 & $0.00$ & 1.40 & $a_1\to \text{CBM}$ & 1.47 & 1.06 & 0.82 & 0.03 \\
$a_1\to e$ & 0.94 & 1.12 & 1.56 & 0.01 & $a_1\to e$ & 0.85 & 1.75 & 1.47 & 0.02\\
$a_1\to e$ & 0.94 & 1.12 & 1.56 & 0.00 &  $a_1\to e$ & 0.84 & 1.82 & 1.45 & 0.01\\
$\text{VBM}\to e$ & 3.04 & 1.66 & 0.22 & 0.01 & $\text{VBM}\to e$ & 2.32 & 4.52 & 0.13 & 0.16\\
$\text{VBM}\to a_1$ & 2.67 & 3.34 & $0.09$ & 0.00 & $\text{VBM}\to a_1$ & 2.61 & 15.70 & 0.05 & 0.08\\

\cline{1-10}

\multicolumn{5}{c}{$\mathbf{kh}$} & \multicolumn{5}{c}{$\mathbf{hk}$}\\
Transition & \makecell{ZPL \\ (eV)} & \makecell{Lifetime \\ ($\mu$s)} & \makecell{$|\vec{\mu}_{xy}|$ \\ (Debye)} & \makecell{$|\vec{\mu}_{z}|$ \\ (Debye)} & Transition & \makecell{ZPL \\ (eV)} & \makecell{Lifetime \\ ($\mu$s)} & \makecell{$|\vec{\mu}_{xy}|$ \\ (Debye)} & \makecell{$|\vec{\mu}_{z}|$ \\ (Debye)}\\
\cline{1-10}
$a_C'\to \text{CBM}$ & 1.43 & 1.87 & 0.64 & 0.13 & $a_C'\to \text{CBM}$ & 1.47 & 1.02 & 0.78 & 0.31 \\
$a_C'\to a_{Si}''$ & 0.83 & 0.33 & 3.47 & 0.40 & $a_C'\to a_{Si}'$ & 0.56 & 0.34 & 4.41 & 4.43\\
$a_C'\to a_{Si}'$ & 0.50 & 2.40 & 1.28 & 1.44 & $a_C'\to a_{Si}''$ & 0.60 & 0.29 & 5.98 & 0.04\\
$\text{VBM}\to a_{Si}'$ & 2.19 & 1.36 & 0.40 & 0.00 & $\text{VBM}\to a_{Si}'$ & 2.27 & 28.3 & 0.08 & 0.00\\
$\text{VBM}\to a_C'$ & 2.20 & 3.01 & 0.27 & 0.00 & $\text{VBM}\to a_C'$ & 2.29 & 35.6 & 0.07 & 0.00\\
\end{tabular}
\end{ruledtabular}
    
    \label{tab:trans}
\end{table*}

As seen in Figs. \ref{fig:hh_bands} and \ref{fig:hk_bands}, the excitations considered in this paper are the lowest possible bound-to-free and free-to-bound transitions as well as excitations between localized states, as indicated by the IPR measure. In this sense, only the smallest transitions from (to) the valence (conduction) band are considered due to the proximity in energy to the known ionization levels. Bound-to-bound transitions involving states inside the valence and conduction bands are in most cases excluded for the same reason or lack of stability in the convergence.

Out of the transitions resulting from these considerations, we make special note of the bound-to-bound transitions in the CAV${}^+$ state denoted by $a_C' \to a_{Si}'$ and $a_C' \to a_{Si}''$, as a possible origin of the AB-lines. These excitations are here seen to create two transitions in each configuration, due to the splitting of the degenerate $e$ states upon occupation. However, it should be noted that there is some difficulty in converging the $a_1 \to e (a_{Si}')$ transition in HSE06, possibly due to the state mixing that could occur between the $a_C'$ and $a_{Si}'$ states because of their similar character during the JT distortion. These particular transitions are therefore in the $hh$ and $kk$ configurations calculated while enforcing the ground state symmetry, thus preventing the JT splitting and resulting in what may be seen as an upper bound for the ZPL. These local transitions can be seen in Table \ref{tab:trans} to be strictly polarized as $E \perp c$ in the $a_1 \to e$ states, while less strictly constrained in the case of $a_C' \to a_{Si}'$ and $a_C' \to a_{Si}''$ transitions in the off-axis configurations, in accordance with selection rules of respective C\textsubscript{3v} and C\textsubscript{1h} symmetries \cite{Castelletto2014}.

These transitions yielded ZPL energies at 0.94, 0.85, 0.83 and 0.60 eV for the $hh$, $kk$, $kh$ and $hk$ respectively, with a respective lifetime for the $a_C' \to a_{Si}''$ of 1.12, 1.75, 0.329 and 0.287 $\mu s$. The calculated ZPL energies thus differ substantially from the AB-lines at 1.8-1.9 eV. The transitions that could match the AB-lines in the CAV\textsuperscript{+} state are the exciton states obtained from $a_C' \to \text{CBM}$ transitions. Such bound exciton transitions can also fit into the proposed working model, given that they exhibit the same group theoretical characteristics the model is based on, giving rise to similar polarization properties and reasonably small separation to the next excited state. These transitions are found to be 1.55, 1.47, 1.47 and 1.43 eV for $hh$, $kk$, $hk$ and $kh$ with respective radiative lifetimes of 0.32, 1.1, 1.0 and 1.9 $\mu$s. The polarization of these transitions are also calculated as mainly $E \perp c$, except in the $hh$ configuration.

One can note that these ZPLs have a significant difference of ${\sim}0.35$ eV from the observed line energy seen in Fig. \ref{fig:AB_figure} at 3.9 K. This may in part be because of finite-size effects caused by the delocalization of the exciton state and insufficient description of the electron-hole interaction. Larger supercell calculations up to $\sim$4600 atoms at PBE level do show a trend of increasing ZPL, but does not indicate a correction much larger than 0.1 eV (see Supplemental Materials \cite{supplemental}). The reader may also note that the excitations discussed thus far have generally shown to have relatively long lifetimes. None of those presented for the positive charge state are likely to agree with the experimentally estimated lifetime of ${\sim}1$ ns \cite{Castelletto2014}.

\begin{figure}[h!]
    \centering
    \includegraphics[width=0.9\columnwidth]{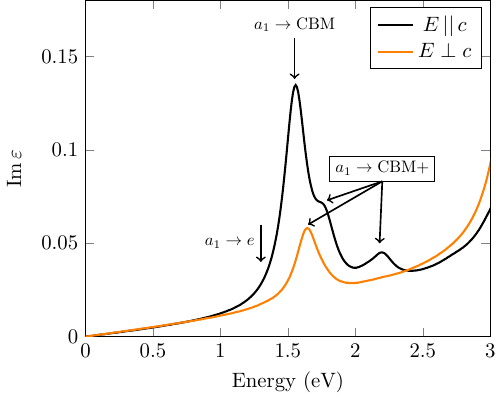}
    \caption{The $G_0W_0$+BSE absorption spectra of the \emph{kk} CAV$+$ state, discerned by either parallel and perpendicular polarization. The y-axis shows the imaginary component of the obtained dielectric function, which is proportional to the absorption coefficient of the system. Arrows point to transitions involving the defect. Note that the defect-local transition $a_1 \to e$ is marked although its oscillator strength is too low to be observed in the spectrum. $a_1 \to \text{CMB}+$ denotes transitions from the defect to above-CBM bands.}
    \label{fig:BSE}
\end{figure}

A GW+BSE calculation on the $kk$ configuration yields the absorption spectrum shown in Fig. \ref{fig:BSE}. The defect-local transition is too weak to be discernible in the spectrum, in agreement with the predicted long lifetimes, but is found at roughly 1.33 eV. The strongest peak in this range is attributed to the lowest bound-to-free transition, found at 1.56 eV, in proximity to the predicted ZPL. We therefore see that the correction of the electron-hole interaction does not significantly increase the ZPL prediction, but rather sets an upper bound on the transition energy still well below the expected AB-line energy.

Interestingly, however, there are three additional contributions appearing at 1.64 eV, 1.77 eV and 2.20 eV, which all involve the excitation to higher conduction band states, revealing possibly complex excitation properties of the defect. The two peaks higher in energy are not as easily dismissed as candidates for the AB-lines based on energy arguments. 

In the neutral charge state, we find bound exciton transitions in $a_{Si}' \to \text{CBM}$ (except for $hk$, with $a_{Si}'' \to \text{CBM}$) in the range of 0.81--0.95 eV as has been predicted in earlier work \cite{Szasz2015}, but here we also report on the radiative lifetimes being on the order of 1 $\mu s$ and above, with a polarization mainly perpendicular to the c-axis with the exception of the $kh$ configuration. For details, see the Supplementary Materials \cite{supplemental}. Other defect transitions do not yield ZPL values that are clearly below the limit of ionization and were also not successfully converged in the $kh$ configuration due to the difficulty in capturing the local states within the conduction band. We do however obtain a 1.16 eV ZPL with a noteworthy radiative lifetime of 30 ns in the $kk$ configuration.

\begin{figure}[h!]
\begin{center}
\includegraphics[width=0.8\columnwidth]{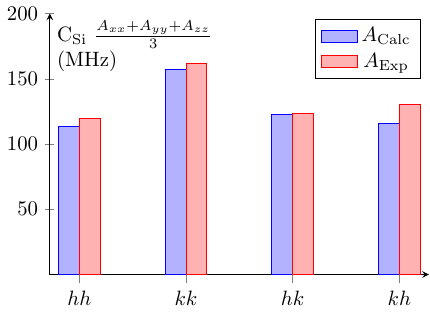}
\caption{Averaged hyperfine parameters of the CAV\textsuperscript{+} state at the C\textsubscript{Si} compared to corresponding experimental findings for the various defect configurations, showing quantitative agreement with experimental findings by Ref.~\onlinecite{Umeda2007}, within an approximately 15\% margin of error.}
\label{fig:cav_hyp_comparison}
\end{center}
\end{figure}
Hyperfine parameters of the antisite carbon are found to have one distinct value along the defect axis, and two nearly identical parameters perpendicular to this axis due to the defect symmetry. These values are in the respective ranges of 117--249 and 26--59 MHz for CAV\textsuperscript{0}, and 245--286 and 57--93 MHz for CAV\textsuperscript{--}, except the $hh$ configuration which displays hyperfine parameters of ${\sim}1$ MHz in the CAV\textsuperscript{--} case. In Fig. \ref{fig:cav_hyp_comparison}, the values for CAV\textsuperscript{+} are presented alongside experimental reference values from Ref.~\onlinecite{Umeda2007}, for which one finds agreement within a 15\% relative margin of error and close to full agreement in the principal axis directions (full disclosure in Supplemental Materials \cite{supplemental}). Hyperfine calculations for the silicon and carbon neighbors yield values in the range of 1--40 MHz.

\begin{table}[h!]
    \centering
    \caption{Zero-field splitting parameters of the neutral CAV triplet ground state.}
    \begin{ruledtabular}
\begin{tabular}{ccr}
    Config. & $D$ (GHz) & $|E|$ (MHz) \\
    \hline
    hh & -1.92 & 568\\
    kk & -2.17 & 684\\
    hk & -1.70 & 12\\
    kh & -2.07 & 356
\end{tabular}
\end{ruledtabular}

    \label{tab:ZFS values}
\end{table}

Only the neutral charge state has more than one unpaired spin in its ground state producing a zero-field splitting. This is calculated to what is shown in Table \ref{tab:ZFS values}. The triplet ground state yields an axial component $D$ in the $\sim-2$ GHz range and a non-zero transversal $E$ component varying within the range of 300--700 MHz with the exception of the $hk$ configuration at an order of 10 MHz.

\section{Discussion}\label{sec:Discussion}

Starting with the ground state properties, Kohn-Sham band structures and wavefunction characteristics are in good agreement with what has previously been reported in other studies of the CAV \cite{Castelletto2014,Umeda2006,Szasz2015}. We find the hyperfine parameters for the positively charged state in excellent agreement with the measurements done by Ref.~\onlinecite{Umeda2007} up to the accuracy expected from HSE06 at $\Gamma$-point \cite{HyperfinePaper}. Due to this both qualitative and quantitative agreement with our results, we are confident that the studied states are the physically formed defect ground states.

Our results also indicate that the CAV defect is able to produce transitions in the 1550 nm (${\sim}0.8$ eV) range suitable for fiber-optical applications. This implies its possible use as a single-photon emitter in quantum communication networks and optics-based quantum information processing\cite{knill2001} and sensing technologies\cite{muller2017}, facilitating the use of existing fiber-optics systems with minimal signal attenuation. Other defects, such as the diamond NV center, require wavelength conversion of their emission to match the desired wavelengths, which can reduce the signal-to-noise ratio by orders of magnitude\cite{dreau2018}. Furthermore, association of the excitation with manipulable and long-lived ground state spin of $S=1$ and $S=1/2$, for the neutral and positive charge state respectively, would make the defect applicable as a spin-photon interface acting as a memory node within quantum networks to allow for long-range transmission with reduced signal loss. 

Starting with the neutral charge state, this was predicted for the bound-to-free excitations calculated in previous works \cite{Szasz2015}. However, our findings on the radiative lifetimes indicate that they may not be particularly bright but sufficiently long-lived for manipulation of the produced exciton state. In contrast, the neutral state of the $kk$ configurations is found to host a localized transition and shows a promisingly small lifetime of just 30 ns at a ZPL of 1.16 eV. However, the ZPL lies on the border of the (0$|$+) ionization threshold \cite{Szasz2015}, and illumination at the necessary wavelength may instead photo-ionize the defect to the positive state and hinder the observation of this emission.

The positive charge state also holds local state transitions that are predicted to fall in the optimal near-infrared range.
The axially symmetric configurations yield ZPLs of 0.85--0.95 eV, although we note that we could not fully relax the \emph{hh} excited state to achieve the correct symmetry. Therefore, the ${\sim}$0.95 eV values may further reduce closer to the ideal 0.8 eV reference. We take special note of the positive $kh$ configuration that shows potential for a $0.83$ eV ZPL in the $a_C' \to a_{Si}''$ transition that also exhibits the lowest radiative lifetime. The lifetime of 0.3 $\mu$s is still considered high in comparison to that of other well-known SiC defects, e.g. the silicon vacancy\cite{nagy2018} or divacancy\cite{falk2014b}, on the order of 10 ns. However, in the context of spin-photon interfaces, even defects such as the T-center in silicon, exhibiting an even longer radiative lifetime (0.94 $\mu$s), have garnered attention as possible spin-photon interfaces by leveraging its emission in the telecom O-band (1326 nm, 0.935 eV), high collection efficiency in the ZPL and its inherent spin dynamics\cite{bergeron2020}. The CAV defect and this specific telecom emission warrant further studies to explore the applicability in quantum optical devices. 

The negative and double positive states do not have luminescence in the near-infrared. This was expected for the negative state due to its ionization threshold of approximately $E_C - 0.5$ eV \cite{Szasz2015}, while we show in this work that the discernibly lowest transitions in the double positive state is $1.6$ eV. These transitions are in the range of the predicted ionization levels or higher, meaning that the charge states are likely limited to these excitations for observable luminescence. One may finally note that the lifetimes calculated for the double positive and negatives states are on the order of 1 $\mu$s or larger, implicating a low brightness.

In regards to the AB-lines, our measurements at low temperatures reveal that the non-thermally activated lines consist of three, rather than four, lines. Furthermore, A2, A3, and A4, together with B1 and B3 must all be high-temperature lines. It is quite certain that B1 and B3 are related to higher excited states of the defects responsible for the appearance of B2 and B4, in agreement with Ref.~\onlinecite{Castelletto2014}. However, it is also possible that the A2, A3 and A4 lines are higher excited states related to B2 and B4, suggesting a more complicated structure of the excited state than the two-level splitting from an $e$ state (in C\textsubscript{3v}) into $a'$ and $a''$ states (in C\textsubscript{1h} symmetry). The currently accepted CAV+ model cannot explain these experimental observations. 

Likewise, there is an absence of predicted transitions in the 1.8--1.9 eV range of the AB spectrum. Transitions between localized defect states are here evaluated to lie below ${\sim}1$ eV, while the bound-to-free transitions, after taking possible supercell-size-effects into account, lie 0.2--0.3 eV below the AB-lines. Improving the accuracy of the electron-hole interaction via $GW$+BSE calculations does not significantly alter this prediction. In light of the calculated radiative lifetimes, it is also uncertain if these transitions could produce the brightness observed at room temperature, on the order of Mcounts per second \cite{Castelletto2014}.

The $GW$+BSE absorption spectrum, however, include several transitions to higher conduction band states, forming possible excitons with an increased likelihood to agree with the AB-line energies, particularly the states found at 1.77 and 2.20 eV in Fig. \ref{fig:BSE}. Attributing the AB-lines to any of these above-CBM states would also be in line with the observed blinking of the AB-line luminescence with low-wavelength lasers \cite{Castelletto2014}, which could then be attributed to photo-ionization in the already delocalized electron-hole state. However, we note that these above-CBM transitions are mainly polarized in $E \parallel c$, and the AB-lines, particularly the B2 and B4 lines, are efficiently excited by $E \perp c$ polarized light. Furthermore, these transitions involve non-degenerate states that should not exhibit a dynamical JT-splitting, and therefore, do not yield an apparent explanation for the appearance of line pairs in the on-axis configurations.

Moreover, considering developments in the optical charge state control of the CAV, measured by electron paramagnetic resonance \cite{Son2019}, our results for the lowest bound-to-free transition are in good agreement with the onset of quenching in the positive state population. In contrast, at illumination energies of the AB-lines ($1.8$ eV) and above, the CAV${}^+$ population is shown to already suffer considerable quenching \cite{Son2019}. 

Unfortunately, one does not find suitable alternatives among the other charge states of this defect. The free-to-bound transitions $\text{VBM} \to a_1 (a_C')$ in the double-positive states might be possible candidates in terms of the ZPL, but fall short when considering the calculated lifetimes, that are even longer than those found for the single-positive state. Considering a different charge state than the positive would also imply abandoning the trends in annealing of the AB-lines observed in earlier works by Steeds \cite{Steeds2002} and more recent confirmations \cite{Karsthof2020}. These associate the disappearance of the lines related to the silicon vacancy due to annealing, with an increase of the AB-lines, which could imply a small energy barrier between the silicon vacancy and the responsible defect, which has been shown for the positive CAV state \cite{Szasz2015}. Our results thus open the model up for new interpretations in terms of the optical properties of the associated defect.

\section{Conclusions}\label{sec:Conclusion}
In this work, we examined the temperature dependence of the AB photoluminescence lines at low temperatures and performed a detailed theoretical characterization of the CAV defect in 4H-SiC, covering optical and spin properties. We found that the AB-line spectrum, in contrast to reports at 80 K, consists of three lines at 649 (A1), 673 (B2) and 677 nm (B4) at 3.9 K, and that the other lines must come from excited states that are thermally activated. Our theoretical investigations on the CAV defect found spin properties, e.g., hyperfine strength and ZFS, in good agreement with experimental findings and predictions in previous works, and revealed optical transitions in the infrared belonging to the CAV${}^+$ and CAV${}^0$ states.

However, our characterization of the CAV defect was unable to conclude a matching set of zero-phonon lines that would suit the AB-lines. In combination with our findings on the low-temperature behavior of the spectrum, the identification of the AB-lines requires reinterpretation and further studies.

While transitions predicted in previous works, such as the bound-to-free transition in the CAV${}^0$ state, are here calculated to have relatively low radiative rates, we find higher rates among the bound-to-bound transitions of the defect. The above-gap defect band of the CAV${}^0$ state features exceptionally low lifetime. However, to ensure its photostability, further studies into the photoionization of the CAV${}^0$ state and the presence of this emission would be of interest.

The CAV${}^+$ is shown to host telecom emission of high interest for quantum optical technologies. This is most relevant for the off-axis configurations where the radiative lifetimes are the lowest and one transition is close to the telecom C-band.

Furthermore, the BSE-spectra calculated in this work indicate several above-CBM excitations available for the CAV${}^+$ state below its predicted ionization energy, which could imply more complex excitation dynamics than previously predicted.

\section*{Acknowledgements}
We acknowledge support from the Knut and Alice Wallenberg Foundation through WBSQD project (Grant No. 2018.0071). I.G.I. acknowledges support from the Swedish Research Council (Grant No. VR 2016-05362). Support from the Swedish Government Strategic Research Area SeRC and the Swedish Government Strategic Research Area in Materials Science on Functional Materials at Linköping University (Faculty Grant SFO-Mat-LiU No. 2009 00971) is gratefully acknowledged. V.I. was supported by the National Research, Development, and Innovation Office of Hungary within the Quantum Information National Laboratory of Hungary (Grant No. 2022-2.1.1-NL-2022-00004) and within grant FK 145395. The computations were enabled by resources provided by the National Academic Infrastructure for Supercomputing in Sweden (NAISS) and the Swedish National Infrastructure for Computing (SNIC) at NSC partially funded by the Swedish Research Council through grant agreements no. 2022-06725 and no. 2018-05973. We acknowledge the EuroHPC Joint Undertaking for awarding project access to the EuroHPC supercomputer LUMI, hosted by CSC (Finland) and the LUMI consortium through a EuroHPC Regular Access call.\\
A.G. acknowledges the National Office of Research, Development, and Innovation of Hungary (NKFIH) Grant No. KKP129866 of the National Excellence Program of Quantum-coherent materials project, the support for the Quantum Information National Laboratory from the Ministry of Culture and Innovation of Hungary (NKFIH Grant No. 2022-2.1.1-NL-2022-00004), the EU H2020 project QuanTELCO (Grant No. 862721) and the EU Horizon project QuMicro (Grant No. 101046911).

\appendix*
\section{Constrained occupation calculations with hybrid functionals using VASP}\label{sec:appendix}
Our theoretical characterization of excited states uses constrained occupation DFT as described in section \ref{sec:methods}. In this approach, one explicitly sets the occupation of the one-electron Kohn-Sham orbitals according to a given orbital order and calculates the corresponding state under this restriction. However, one must ensure that the desired states remain occupied throughout the calculation. Hence, it is necessary to prevent the reordering of orbitals when seeking the excited state electron density.

Excited state calculations with hybrid functionals are particularly prone to reordering the orbitals since the exact-exchange term generally lowers the energy of occupied states and increases it for unoccupied states. If the orbitals are allowed to reorder, the calculation may either not converge or end in an unintended final state. This can also happen if one does not have a pre-defined orbital ordering at the start of the calculation. It is therefore preferable to start from a pre-created WAVECAR and turn off explicit eigenstate diagonalization with LDIAG=FALSE.

In calculations using the VASP software, the FERWE and FERDO input parameters restrict the orbitals to occupy. However, for such calculations, the resulting behavior of the input parameters needed to keep the orbitals from reordering differ between versions of VASP. In VASP 5.4.1 (and possibly earlier versions), if the calculation is started with a pre-created WAVECAR file (ISTART=1) and preventing statespace diagonalization, the desired result is achieved when using the damped velocity friction algorithm (ALGO=Damped). However, the use of other algorithms changes the behavior of the subspace rotations for this kind of calculation, which prevents it from converging into the desired final state. Furthermore, later VASP versions (5.4.4, 6.2.0 and 6.2.1) change the settings for ALGO=Damped so that the subspace rotations behave the same way as for the other ALGO settings.

We have found how to modify the VASP source code so that the change in subspace rotations can be avoided under these conditions for all relevant ALGO choices and earlier versions. We are happy to share these changes with others working with similar calculations in these earlier versions. After reaching out to the VASP developers on this matter, the corresponding change was implemented as of version 6.3.0.


%

\end{document}


\title{Temperature dependence of the AB-lines and Optical Properties of the Carbon-Antisite Vacancy Pair in 4H-SiC\\Supplementary Information}
\author{Oscar Bulancea-Lindvall}%
\author{Joel Davidsson}%
\author{Ivan G. Ivanov}%
\affiliation{Department of Physics, Chemistry and Biology (IFM), Linköping University, SE-58183 Linköping, Sweden}
\author{Adam Gali}%
\affiliation{HUN-REN Wigner Research Centre for Physics, P.O. Box 49, H-1525 Budapest, Hungary}
\affiliation{Department of Atomic Physics, Institute of Physics, Budapest University of Technology and Economics, Műegyetem rakpart 3., H-1111 Budapest, Hungary}
\affiliation{MTA-WFK Lend\"ulet "Momentum" Semiconductor Nanostructures Research Group}
\author{Viktor Ivády}%
\affiliation{Department of Physics, Chemistry and Biology (IFM), Linköping University, SE-58183 Linköping, Sweden}
\affiliation{Department of Physics of Complex Systems, E\"otv\"os Loránd University, Egyetem tér 1-3, H-1053, Budapest, Hungary}
\affiliation{MTA–ELTE Lend\"{u}let "Momentum" NewQubit Research Group, Pázmány Péter, Sétány 1/A, 1117 Budapest, Hungary}
\author{Rickard Armiento}%
\author{Igor A. Abrikosov}%
\affiliation{Department of Physics, Chemistry and Biology (IFM), Linköping University, SE-58183 Linköping, Sweden}

\maketitle

\newpage

\section{Detailed account of CAV optical characterization}

\begin{table*}[ht!]
    \renewcommand{\arraystretch}{1.3}
    \centering
    \caption{Properties of the studied excitations, showing ZPL, radiative lifetime and dipole moment as calculated from first principles, categorized by the exhibiting ground state. Transitions are labeled by the resulting hole and occupied state, discerning between the two $a'$ states formed from the $a_1$ and $e$ states under symmetry breaking by the localization around the carbon ($a_1 = a_C'$) or the vacancy-neighboring silicon ($e$, or $a_{Si}'$ and $a_{Si}''$). Noteworthy are the bound-to-bound transitions in the positive state and bound-to-free transitions in the neutral state of which most are in the near-infrared.}
    \begin{ruledtabular}
\begin{tabular}{cccccc|cccccc}
\multicolumn{6}{c}{$\mathbf{hh}$} & \multicolumn{6}{c}{$\mathbf{kk}$}\\
State & Transition & \makecell{ZPL \\ (eV)} & \makecell{Lifetime \\ ($\mu$s)} & \makecell{$|\vec{\mu}_{xy}|$ \\ (Debye)} & \makecell{$|\vec{\mu}_{z}|$ \\ (Debye)} & State & Transition & \makecell{ZPL \\ (eV)} & \makecell{Lifetime \\ ($\mu$s)} & \makecell{$|\vec{\mu}_{xy}|$ \\ (Debye)} & \makecell{$|\vec{\mu}_{z}|$ \\ (Debye)}\\
\cline{1-12}
\multirow{1}{*}{CAV\textsuperscript{++}} & $\text{VBM}\to a_1$ & 1.60 & 3.94 & 0.38 & 0.00 & \multirow{1}{*}{CAV\textsuperscript{++}} & $\text{VBM}\to a_1$ & 1.64 & 2.89 & 0.33 & 0.26\\
\cline{1-12}
\multirow{5}{*}{CAV\textsuperscript{+}} & $a_1\to \text{CBM}$ & 1.55 & 0.32 & $0.00$ & 1.40 & \multirow{5}{*}{CAV\textsuperscript{+}} & $a_1\to \text{CBM}$ & 1.47 & 1.06 & 0.82 & 0.03 \\
&$a_1\to e$ & 0.94 & 1.12 & 1.56 & 0.01 & &$a_1\to e$ & 0.85 & 1.75 & 1.47 & 0.02\\
&$a_1\to e$ & 0.94 & 1.12 & 1.56 & 0.00 &  &$a_1\to e$ & 0.84 & 1.82 & 1.45 & 0.01\\
&$\text{VBM}\to e$ & 3.04 & 1.66 & 0.22 & 0.01 & &$\text{VBM}\to e$ & 2.32 & 4.52 & 0.13 & 0.16\\
&$\text{VBM}\to a_1$ & 2.67 & 3.34 & $0.09$ & 0.00 & &$\text{VBM}\to a_1$ & 2.61 & 15.70 & 0.05 & 0.08\\
\cline{1-12}
\multirow{3}{*}{CAV\textsuperscript{0}} & $a_{Si}'\to \text{CBM}$ & 0.81 & 7.77 & 0.53 & 0.04 & \multirow{3}{*}{CAV\textsuperscript{0}} & $a_{Si}'\to \text{CBM}$ & 0.83 & 1.26 & 1.79 & 0.22\\
 & $a_C'\to a_{Si}''$ & 1.42 & 0.33 & 1.55 & 0.01 &  & $a_C'\to a_{Si}''$ & 1.16 & 0.028 & 4.67 & 5.56\\
 & $\text{VBM}\to a_C'$ & 2.69 & 1.28 & 0.28 & 0.12 &  & $\text{VBM}\to a_C'$ & 2.70 & 8.66 & 0.10 & 0.05\\
\cline{1-12}
\multirow{2}{*}{CAV\textsuperscript{-}} & $a_{Si}'\to \text{CBM}$ & 0.64 & 11.3 & 0.88 & 0.00 & \multirow{2}{*}{CAV\textsuperscript{-}} & $a_{Si}'\to \text{CBM}$ & 0.45 & 15.47 & 1.29 & 0.00\\
 & $a_C'\to \text{CBM}$ & 0.44 & 41.8 & 0.80 & 0.00 &  & $a_C'\to \text{CBM}$ & 0.43 & 10.03 & 1.68 & 0.00\\

\cline{1-12}

\multicolumn{6}{c}{$\mathbf{kh}$} & \multicolumn{6}{c}{$\mathbf{hk}$}\\
State & Transition & \makecell{ZPL \\ (eV)} & \makecell{Lifetime \\ ($\mu$s)} & \makecell{$|\vec{\mu}_{xy}|$ \\ (Debye)} & \makecell{$|\vec{\mu}_{z}|$ \\ (Debye)} & State & Transition & \makecell{ZPL \\ (eV)} & \makecell{Lifetime \\ ($\mu$s)} & \makecell{$|\vec{\mu}_{xy}|$ \\ (Debye)} & \makecell{$|\vec{\mu}_{z}|$ \\ (Debye)}\\
\cline{1-12}
\multirow{1}{*}{CAV\textsuperscript{++}} & $\text{VBM}\to a_C'$ & 1.68 & 0.68 & 0.85 & 0.00 & \multirow{1}{*}{CAV\textsuperscript{++}} & $\text{VBM}\to a_C'$ & 1.64 & 49.6 & 0.10 & 0.01\\
\cline{1-12}
\multirow{5}{*}{CAV\textsuperscript{+}} & $a_C'\to \text{CBM}$ & 1.43 & 1.87 & 0.64 & 0.13 & \multirow{5}{*}{CAV\textsuperscript{+}} & $a_C'\to \text{CBM}$ & 1.47 & 1.02 & 0.78 & 0.31 \\
&$a_C'\to a_{Si}''$ & 0.83 & 0.33 & 3.47 & 0.40 & &$a_C'\to a_{Si}'$ & 0.56 & 0.34 & 4.41 & 4.43\\
&$a_C'\to a_{Si}'$ & 0.50 & 2.40 & 1.28 & 1.44 & &$a_C'\to a_{Si}''$ & 0.60 & 0.29 & 5.98 & 0.04\\
&$\text{VBM}\to a_{Si}'$ & 2.19 & 1.36 & 0.40 & 0.00 & &$\text{VBM}\to a_{Si}'$ & 2.27 & 28.3 & 0.08 & 0.00\\
&$\text{VBM}\to a_C'$ & 2.20 & 3.01 & 0.27 & 0.00 & &$\text{VBM}\to a_C'$ & 2.29 & 35.6 & 0.07 & 0.00\\
\cline{1-12}
\begin{tabular}{c}
CAV\textsuperscript{0}
\end{tabular}
&
\begin{tabular}{c}
$a_{Si}'\to \text{CBM}$\\
$\text{VBM}\to a_C'$
\end{tabular}
&
\begin{tabular}{c}
0.94\\
2.55
\end{tabular}
&
\begin{tabular}{c}
1.19\\
2.16
\end{tabular}
&
\begin{tabular}{c}
1.03\\
0.22
\end{tabular}
&
\begin{tabular}{c}
1.12\\
0.13
\end{tabular}
&
\begin{tabular}{c}
CAV\textsuperscript{0}
\end{tabular}
&
\begin{tabular}{c}
$a_{Si}''\to \text{CBM}$\\
$a_C'\to a_{Si}'$\\
$\text{VBM}\to a_{Si}''$
\end{tabular}
&
\begin{tabular}{c}
0.82\\
1.03\\
2.53
\end{tabular}
&
\begin{tabular}{c}
1.28\\
1.69\\
4.06
\end{tabular}
&
\begin{tabular}{c}
1.79\\
1.11\\
0.17
\end{tabular}
&
\begin{tabular}{c}
0.00\\
0.00\\
0.09
\end{tabular}\\
\cline{1-12}
\multirow{2}{*}{CAV\textsuperscript{-}} & $a_C'\to \text{CBM}$ & 0.59 & 2.61 & 2.1 & 0.00 & \multirow{2}{*}{CAV\textsuperscript{-}} & $a_C'\to \text{CBM}$ & 0.48 & 72.77 & 0.53 & 0.00\\
 & $a_{Si}'\to \text{CBM}$ & 0.60 & 1.80 & 1.67 & 1.81 &  & $a_{Si}'\to \text{CBM}$ & 0.51 & 12.70 & 0.36 & 1.10\\
\end{tabular}
\end{ruledtabular}
    
    \label{tab:trans}
\end{table*}

\section{Hyperfine parameters}

\begin{table*}[hbt!]
\centering
\caption{Hyperfine strength of the antisite carbon, C\textsubscript{Si}, in the CAV\textsuperscript{+} state, shown together with their average and the angle between strongest principal axis and c-axis of the bulk in comparison with values measured by Umeda et al.\cite{Umeda2007}}
\begin{ruledtabular}

\begin{tabular}{cccccc}
 & \thead{$A_{xx}$ (Exp) \\ MHz } & \thead{$A_{yy}$ (Exp) \\ MHz } & \thead{$A_{zz}$ (Exp) \\ MHz} & \thead{$(A_{xx} + A_{yy} + A_{zz})/3$  \\ (Exp) MHz} &
 $\theta_{z\cdot c}$ (Exp)\\
 \hline
 hh & 53.0 (63.5) &	53.0 (63.5) & 235 (231) & 114 (119)  & 0.00${}^\circ$ (0$^\circ$)\\
 kk & 92.0 (104) &	92.0 (104) & 289 (279) & 158 (162) & 0.00${}^\circ$ (0$^\circ$)\\
 hk & 59.6 (64.7) &	59.7 (68.6) & 249 (236) & 123 (123) & 108.8$^\circ$ (109${}^\circ$)\\
 kh & 54.1 (72.8) & 54.1 (74.2) & 240 (245) & 116 (131)  & 109.4$^\circ$ (110${}^\circ$)
\end{tabular}
\end{ruledtabular}
\label{tab:cav_plus_hyp}
\end{table*}

In Table \ref{tab:cav_plus_hyp}, we see general agreement between calculated hyperfine elements and experimental observation in the literature. Overall, there is at most a 20\% discrepancy between theory and experiment for any single value. The direction of the eigenvector $z$-axis is also well-aligned with the measured orientation, placing along the symmetry axis or within the plane of reflection in the C\textsubscript{3v} and C\textsubscript{1h} cases, respectively.

\section{Finite-size effects of the defect-to-band transition in the CAV+ state}

\begin{figure}[h!]
    \centering
    \includegraphics{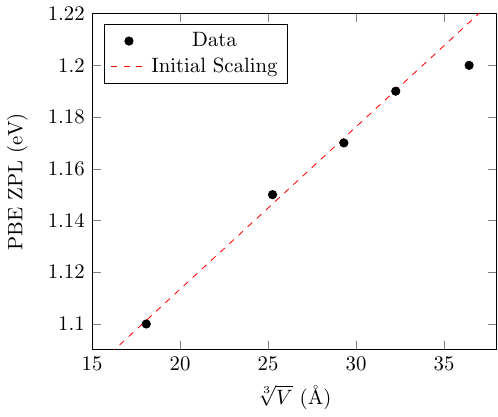}
    \caption{The ZPL of the $a_1 \to \text{CBM}$ transition in the \emph{hh} configuration, evaluated with $\Gamma$-point PBE for supercell sizes ranging from 576 atom to ca 4600 atoms. The dashed line illustrates the linear scaling in effective supercell side length seen for smaller sizes.}
    \label{fig:ZPL_scaling}
\end{figure}

Scaling of the $a_1 \to \text{CBM}$ ZPL in the CAV+ \emph{hh} configuration was tested on PBE-structures calculated at $\Gamma$-point, with results shown in Fig. \ref{fig:ZPL_scaling}. From the original structure used throughput the paper, the 576 atom supercell, to the scaled structure containing ca 3600 atoms, the ZPL displays linear scaling owing the continued extension of the defect excited state wavefunction and the effect of charged periodic image interactions within the applied charge-jellium. The scaling deviates first at a structure size of ca 4600 atoms, with an effective side length of 36.4 \AA\ and ZPL of 1.2 eV, which is assumed to mark the start of final convergence. From this, we conclude that finite-size effects in the CAV+ state accounts for roughly a 0.1 eV decrease in ZPL energy, assuming the scaling is independent of details in the lattice structure.

\section{$GW$+BSE spectrum of the CAV+ state}

\begin{figure}[h!]
    \centering
    \begin{minipage}{0.48\textwidth}
    \tikz{
        \node at (0,0) {\includegraphics[width=\textwidth]{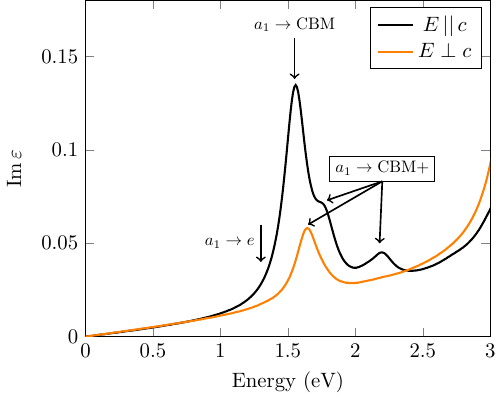}};
        \node [scale=1.4] at ({-0.4*\textwidth},3.7) {a)};
        \node [scale=1.4] at ({-0.2*\textwidth},2.6) {\emph{kk}};
    }
    \end{minipage}
    \quad
    \begin{minipage}{0.48\textwidth}
    \tikz{
        \node at (0,0) {\includegraphics[width=\textwidth]{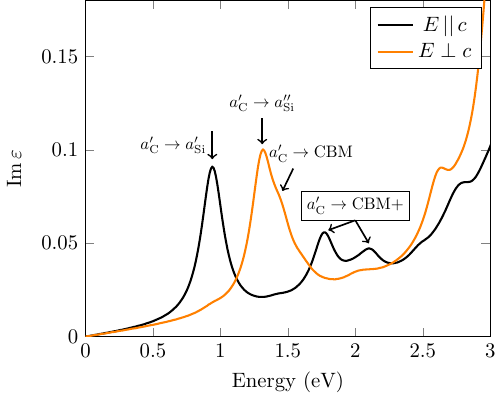}};
        \node [scale=1.4] at ({-0.4*\textwidth},3.7) {b)};
        \node [scale=1.4] at ({-0.2*\textwidth},2.6) {\emph{kh}};
    }
    \end{minipage}
    \begin{minipage}{0.48\textwidth}
    \tikz{
        \node at (0,0) {\includegraphics[width=\textwidth]{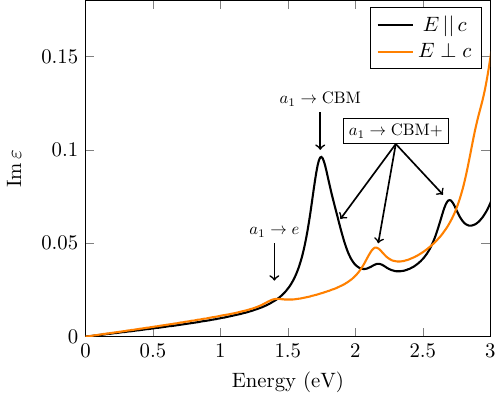}};
        \node [scale=1.4] at ({-0.4*\textwidth},3.7) {c)};
        \node [scale=1.4] at ({-0.2*\textwidth},2.6) {\emph{hh}};
    }
    \end{minipage}
    \quad
    \begin{minipage}{0.48\textwidth}
    \tikz{
        \node at (0,0) {\includegraphics[width=\textwidth]{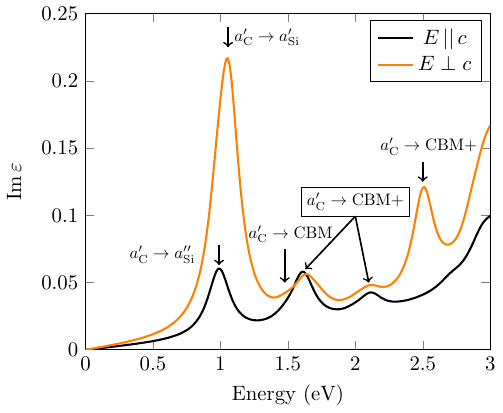}};
        \node [scale=1.4] at ({-0.4*\textwidth},3.7) {d)};
        \node [scale=1.4] at ({-0.2*\textwidth},2.6) {\emph{hk}};
    }
    \end{minipage}
    \caption{$GW$+BSE spectra of all the CAV configurations in the positive ground state. }
    \label{fig:GW_BSE}
\end{figure}

\todo[inline]{Mention what has been written about \emph{kk} in paper}
The \emph{kk} configuration in Fig. \ref{fig:GW_BSE}a) is discussed in the main text, and is representative of the on-axis structures, although the \emph{hh} spectrum CBM-line is shifted to higher energies. However, the \emph{hh} configuration in Fig. \ref{fig:GW_BSE}c) differs in the polarization of the lowest defect-to-band transition. The absorption of $a_{\text{C}}'\to a_{\text{Si}}'$ and $a_{\text{C}}'\to a_{\text{Si}}'$ in the \emph{kh} configuration are seen to differ by ca 0.35 eV, which reflects the predicted difference in Fig. \ref{tab:trans} while also agreeing with the predicted polarization, albeit not indicating the same relation among the predicted lifetimes. Comparing the \emph{kh} and \emph{kk} spectra, it is clear that the $a_1(a_C') \to \text{CBM}$ is considerably weaker for the off-axis configurations, which also agrees with the characterization at the DFT-level of theory. We also note that all spectra contain above-CBM exciton states with absorption at ca 1.6-1.8 eV and 2.2 eV, although with varying amplitudes and polarizations. 

Regarding a match for the AB-lines, one notes the absence in \emph{kk} and \emph{kh} of significant $E\perp c$ polarized lines in the AB-line range. The nearly indiscernible peaks in the \emph{kh} configuration at roughly 2.05 eV together with the neighboring above-CBM state makes for an interesting suggestion for one AB-line pair, such as the B1 and B2 which match in polarization. Similar appearances in the \emph{hh} and \emph{hk} spectra are different in nature, as the \emph{hk} peaks in each polarization is believed to be from the same transition, while the \emph{hh} $E \perp c$ peak hosts two transitions to degenerate states. The \emph{hh} peak could therefore host the B3 and B4 lines, which are mainly $E \perp c$-polarized, while the neighboring $E \parallel c$ above-CBM peak could be attributed to a thermally-activated line. However, in all proposed cases, the amplitudes are exceedingly small compared to other present transitions that have neither been predicted as bright, nor previously been identified with the CAV defect. In addition, assuming the above-CBM peaks to be responsible for the observed lines would also imply quite complex optical properties, with a balance between photo-ionization and optical excitation to the various excitons predicted in Fig. \ref{fig:GW_BSE}. 


%
